\documentclass[aps,pra,twocolumn,superscriptaddress,epsfig]{revtex4-1}

\usepackage{amsmath}
\usepackage{amsfonts}
\usepackage{amssymb}
\usepackage{amstext}
\usepackage{amsthm}
\usepackage{graphicx}
\usepackage{graphics}
\usepackage{times}
\usepackage{dcolumn}
\usepackage{bm}
\usepackage{color}
\usepackage{latexsym}
\usepackage{hyperref}
\usepackage{subfigure}
\usepackage[utf8]{inputenc}
\usepackage{textcomp}
\usepackage{dsfont}
 
\def\bra#1{\langle{#1}\vert}
\def\ket#1{\vert{#1}\rangle}
\def\braket#1{\langle{#1}\rangle}

{\catcode`\|=\active 
 \gdef\Braket#1{\begingroup
\mathcode`\|32768\let\vert\BraVert\left<{#1}\right>\endgroup}}
\def\BraVert{\egroup\,\mid\,\bgroup}

\def\tr{\mbox{tr}}
\newcommand{\ent}{\mathbf{S}}
\newcommand{\q}{\mathbf{Q}}
\newcommand{\anc}{\mathcal{A}}
\newcommand{\s}{\mathcal{S}}
\newcommand{\e}{\mathcal{R}}
\newcommand{\se}{\mathcal{RS}}

\newcommand{\z}{\mathcal{Z}}

\def\d{{\rm d}}

\begin{document}

\title{Experimental demonstration of information to energy conversion in a quantum system at the Landauer Limit}

\author{J. P. S. Peterson}
\affiliation{Centro Brasileiro de Pesquisas F\'{i}sicas, Rua Dr. Xavier Sigaud 150, 22290-180 Rio de Janeiro, Brasil}

\author{R. S. Sarthour}
\affiliation{Centro Brasileiro de Pesquisas F\'{i}sicas, Rua Dr. Xavier Sigaud 150, 22290-180 Rio de Janeiro, Brasil}

\author{A. M. Souza}
\affiliation{Centro Brasileiro de Pesquisas F\'{i}sicas, Rua Dr. Xavier Sigaud 150, 22290-180 Rio de Janeiro, Brasil}

\author{I. S. Oliveira}
\affiliation{Centro Brasileiro de Pesquisas F\'{i}sicas, Rua Dr. Xavier Sigaud 150, 22290-180 Rio de Janeiro, Brasil}

\author{J. Goold}
\affiliation{The Abdus Salam International Centre for Theoretical Physics (ICTP), Trieste, Italy}

\author{K. Modi}
\affiliation{School of Physics and Astronomy, Monash University, Victoria 3800, Australia}

\author{D. O. Soares-Pinto}
\affiliation{Instituto de F\'{i}sica de S\~{a}o Carlos, Universidade de S\~{a}o Paulo, CP 369, 13560-970, S\~{a}o Carlos, SP, Brasil}

\author{L. C. C\'{e}leri}
\affiliation{Instituto de F\'{i}sica, Universidade Federal de Goi\'{a}s, Caixa Postal 131 74001-970, Goi\^{a}nia, Brasil}

\begin{abstract}
Landauer's principle sets fundamental thermodynamical constraints for classical and quantum information processing, thus affecting not only various branches of physics, but also of computer science and engineering. Despite its importance, this principle was only recently experimentally considered for classical systems. Here we employ a nuclear magnetic resonance setup to experimentally address the information to energy conversion in a quantum system. Specifically, we consider a three nuclear spins $S=1/2$ (qubits) molecule ---the system, the reservoir and the ancilla--- to measure the heat dissipated during the implementation of a global system-reservoir unitary interaction that changes the information content of the system. By employing an interferometric technique we were able to reconstruct the heat distribution associated with the unitary interaction. Then, through quantum state tomography, we measured the relative change in the entropy of the system. In this way we were able to verify that an operation that changes the information content of the system must necessary generate heat in the reservoir, exactly as predicted by Landauer's principle. The scheme presented here allows for the detailed study of irreversible entropy production in quantum information processors.
\end{abstract}

\maketitle

\section{Introduction}

In 1961, Rolf Landauer demonstrated a revolutionary principle which provided a definitive link between the information theory and thermodynamics \cite{Landauer}. Landauer's principle states that in any irreversible computation there is an unavoidable entropy production, manifested as heat, which is dissipated to the non-information bearing degrees of freedom of the computer. Landauer discovered that this dissipated heat is bounded, from below, by the information theoretical entropy change. Some years later Charles Bennett \cite{Bennett} and independently Oliver Penrose \cite{Penrose} used this principle to explain how to solve the long standing Maxwell's demon problem in thermodynamics. The demon as first conceived by Maxwell \cite{Maxwell}, and named by Kelvin \cite{Kelvin}, has had an infamous and often controversial history which spans the entire development of statistical mechanics \cite{Szilard, Leff, Nori}. Controversies and philosophical issues aside, both the demon and Landauer's principle have, at their core, simple but pragmatic applications. Landauer's principle sets fundamental thermodynamic constraints for (classical and quantum) information processing.

Almost half a century has passed and the Landauer limit has finally been reached in several experiments on classical platforms \cite{Orlov, udea, Lutz, Koski, Jun}. This delay is due to the fundamental difficulty of dealing with systems containing only a few degrees of freedom, where the fluctuations about average behavior are dominant. For these systems the concept of large numbers and hence any notion of thermodynamic equilibrium does not hold. However, the past 20 years we have seen a rapid progress in non-equilibrium statistical mechanics with the development of stochastic thermodynamics \cite{Sekimoto:10} and the associated discovery of various fluctuation theorems \cite{Seifert}. Within this framework, thermodynamic quantities such as heat, work, and entropy now become stochastic variables described by appropriate probability distributions over individual phase space trajectories. This approach not only allows physicists to explore the ultimate thermodynamic limits of microscopic systems but also their information processing capabilities \cite{Orlov, udea, Lutz, Koski, Jun}.

Turning towards quantum systems \cite{Goold:rev, Sai:rev}, a picture of non-equilibrium thermodynamics has also emerged with thermodynamic quantities such as heat, work and entropy, being formulated as stochastic variables \cite{esposito, mrev}. As expected, in the quantum domain, the situation is even harder. The absence of a phase space picture due to intrinsic quantum uncertainty aside, one also has to cope with the necessity of performing invasive projective measurements on to a time dependent energy eigenbasis \cite{lutz}. Until recently, this restrictive necessity has hindered experimental advances in studying the non-equilibrium thermodynamics of quantum systems. However, recent proposals have outlined that the quantum work distribution maybe extracted without the need of performing these direct measurements in favour of implementing phase estimation of an appropriately coupled ancilla \cite{Dorner, Mazzola}, which samples the characteristic function of the distribution of the thermodynamic quantity. These schemes were recently implemented experimentally and allowed for the first verification of the quantum work fluctuation relations on a Nuclear Magnetic Resonance (NMR) system \cite{Batalhao}.

In this work we used a nuclear magnetic resonance setup to measure the heat dissipated in elementary quantum logic gates at the Landauer limit. Specifically, we consider a three qubit sample ---the system, the reservoir and the ancilla--- in order to measure the heat dissipated during the implementation of a global system-reservoir unitary that changes the information content of the system. We do this in two independent steps. First, by employing an interferometric technique, using the ancilla, we were able to reconstruct the heat distribution associated with the unitary process. Secondly, through quantum state tomography we measure the change in the entropy of the system. In this way we were able to verify that an operation that changes the information content of the system must necessary generate heat in the reservoir, exactly as predicted by Landauer's principle. The protocol used in this work allows the detailed study of irreversible entropy production in quantum information processors. 

\begin{figure}[t]
\begin{center}
\includegraphics[scale=0.25]{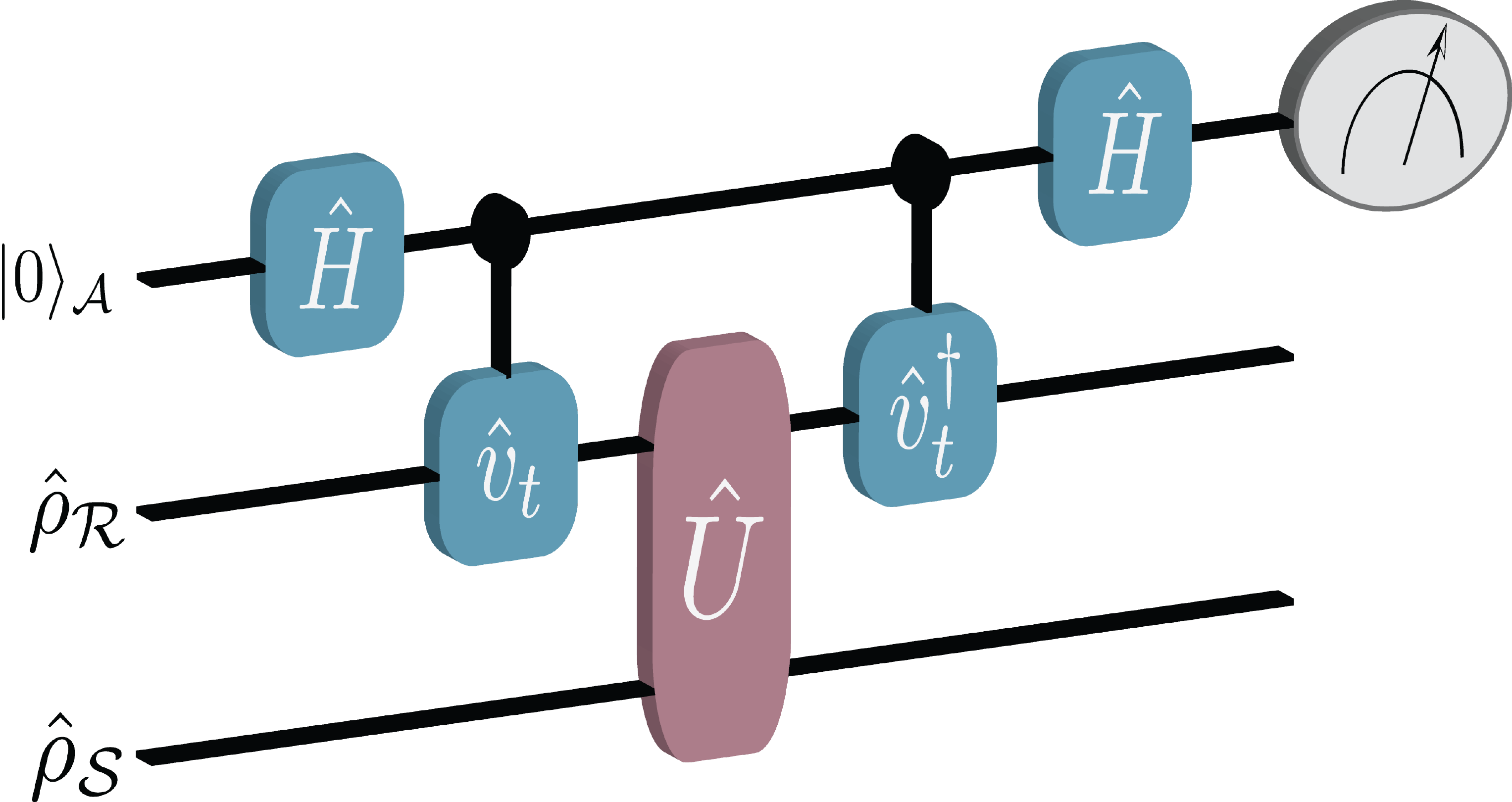}
\end{center}
\caption{{\bf Quantum circuit.} As outlined in \cite{John} by employing the ancilla system we perform the appropriate phase estimation as illustrated in panel. The Hadamard gates $H$ and the controlled operations $\hat v_t = \exp\{-i \hat H_{\e}t\}$ and its Hermitian conjugate $\hat v^{\dagger}_t$ are necessary to effectively make an interferometer. By varying the time $t$ of this operation the associated phase difference can be measured and, by appropriate measurements of the ancilla, we can reconstruct the characteristic function, given in Eq.~\eqref{charfunction}, of the distribution of heat dissipated to the reservoir.}
\label{fig:circ}
\end{figure}

\section{Landauer processes and heat Statistics}
Imagine a quantum system $\s$ in state $\hat \rho_\s$ containing some information that we want to erase. We can then send the system through an erasure machine that reduces the amount of information in the system. Landauer's principle relates the heat generated in the erasure machine to the change in the information entropy of the system. This relationship holds as
\begin{gather}\label{landauer}
\beta \braket{\q} \ge \Delta \ent,
\end{gather}
where $\braket{\q}$ is the average heat dissipated to the reservoir and $\Delta \ent = \ent_i - \ent_f$ is the change of von Neumann entropy. The von Neumann entropy is a measure of the information contained in a quantum state $\hat \rho_\s$ as $\ent (\hat{\rho}) = -\tr [\hat{\rho} \log(\hat{\rho})]$. We will formally define $\q$ below.

The erasure machine performs a generalised quantum operation on $\s$, thus it can be represented by a completely positive trace preserving (CPTP) map. Any CPTP map can be seen as contraction of unitary dynamics of $\s$ along with a reservoir $\e$ \cite{Nielsen}. Recently Reeb \& Wolf have shown that Eq.~\eqref{landauer} holds for processes satisfying the following criteria \cite{reeb}: (i) The process involves a system $\s$ and a reservoir $\e$; (ii) The initial $\se$ state is uncorrelated, i.e., $\hat\rho_{\se} = \hat\rho_\s \otimes \hat\rho_\e$; (iii) The reservoir $\e$ is initially in the Gibbs state $\hat\rho_\e = \exp\{-\beta \hat H_\e\}/\z_\e$, with Hamiltonian $\hat H_\e = \sum_m E_m \ket{r_m}\bra{r_m}$, inverse temperature $\beta^{-1} = k_B T$\footnote{For simplicity, from here on we set Boltzmann constant $k_B =1$.}, and partition function $\z_\e =\tr[\exp\{-\beta \hat H_\e\}]$; (iv) The interaction between $\s$ and $\e$ is unitary: $\hat\rho'_{\se} = \hat U \hat\rho_{\se} \hat U^\dag$. 

Relaxing any of these four criteria can lead to violations of Eq.~\eqref{landauer}. Thus we call processes satisfying the four criteria above as \emph{Landauer processes}. The resultant dynamics on the system or the reservoir alone is non-unitary. This is responsible for generating heat in $\e$ at the expense of changing the entropy of $\s$. The change in entropy of $\s$ can be computed by calculating the entropies of $\hat \rho_\s$ and $\hat \rho'_\s$. While the average heat on the reservoir is given as
\begin{gather}\label{avgheat}
\braket{\q} = \tr[\hat H_\e (\hat \rho'_\e - \hat \rho_\e)].
\end{gather}
Here $\hat \rho'_\s = \tr_\e [\hat \rho'_\se]$ and $\hat \rho'_\e = \tr_\s [\hat \rho'_\se]$. While the entropy change of $\s$ can be computed by measuring the states of $\s$ before and after the Landauer process, measuring heat is not so straightforward.

Moreover, $\q$ is a stochastic variable. That is, in a given run the reservoir may be in state $\ket{r_m}$, which has energy $E_m$, with probability $p_m = \exp\{-\beta \hat E_m \}/\z_\e$. After the process we may find the reservoir in state $\ket{r_n}$ with energy $E_n$. The probability for finding $\e$ in state $\ket{r_n}$ is given by
\begin{gather}
p_{n|m} = \tr[\hat U \ket{r_{m}} \bra{r_{m}} \otimes \hat\rho_\s \, \hat U^{\dagger} \ket{r_{n}} \bra{r_{n}}].
\end{gather}
The last equation comes from criteria (iv) above when the initial $\e$ state is set to $\ket{r_n}$. Then with probability $p_{n|m} p_m$ the reservoir has $E_n - E_m$ heat generation. These probabilities give us a distribution for the heat \cite{GPM, taranto}
\begin{gather}
\label{heatdistribution} 
P(\q)=\sum_{mn}p_{m}p_{n|m} \delta(\q-(E_{n}-E_{m})).
\end{gather}
The first moment of this distribution, $\braket{\q}=\int P(\q)\q \d\q$, is exactly the average heat of Eq.~\eqref{avgheat}.

If we can measure the entire heat distribution we can measure the average heat. However, due to the invasive nature of projective measurements, it is generally not easy to measure the heat distribution, rather we measure its corresponding characteristic function, $\Theta(t)$, calculated by Fourier transform to be
\begin{eqnarray}
\Theta(t) &=& \sum_{mn} p_{m} p_{n|m} e^{-i(E_{n}-E_{m})t} \nonumber \\
&=& \tr[\hat U \, \hat\rho_\e \, \hat v_t^{\dagger} \otimes \hat\rho_\s \, \hat U^{\dagger} \, \hat v_t], 
\label{charfunction}
\end{eqnarray}
where $\hat v_t=e^{i \hat H_\e t}$ is a unitary transformation on $\e$. The details for relating the characteristic function to work and heat distributions can be found in \cite{Dorner, Mazzola, John}.

In order to measure $\Theta(t)$ we implemented the circuit shown in Fig.~\ref{fig:circ}. In this method we have utilised an ancillary qubit (labelled as $\anc$) in the superposition state $\ket{+} = (\ket{0} +\ket{1}) /\sqrt{2}$. The implementation of unitary operations $\hat v_t$ and $\hat v_t^\dag$ is controlled by the state of $\anc$; the operations are applied when $\anc$ is in state $\ket{1}$ and not applied when the state is $\ket{0}$. Between the two controlled operations, the system and the reservoir interact via $\hat U$. The expectation values for observable $\hat{\sigma}_{x}$ and $\hat{\sigma}_{y}$ on $\anc$ are directly related to the characteristic function $\Theta(t) = \braket{ \hat{\sigma}_{x}(t)}_{\anc } - i \braket{ \hat{\sigma}_{y}(t)}_{\anc }$. In other words, we employed an interferometric technique to map the information about the characteristic function of the desired heat distribution, or the Fourier transform of it, on the state of the ancilla \cite{Dorner, Mazzola, John}. Next, we present our experimental setup to observe the information to energy conversion of basic quantum logic gates which can be studied in the quantum domain using a NMR system.

\begin{figure}[t]
\begin{center}
\includegraphics[scale=0.25]{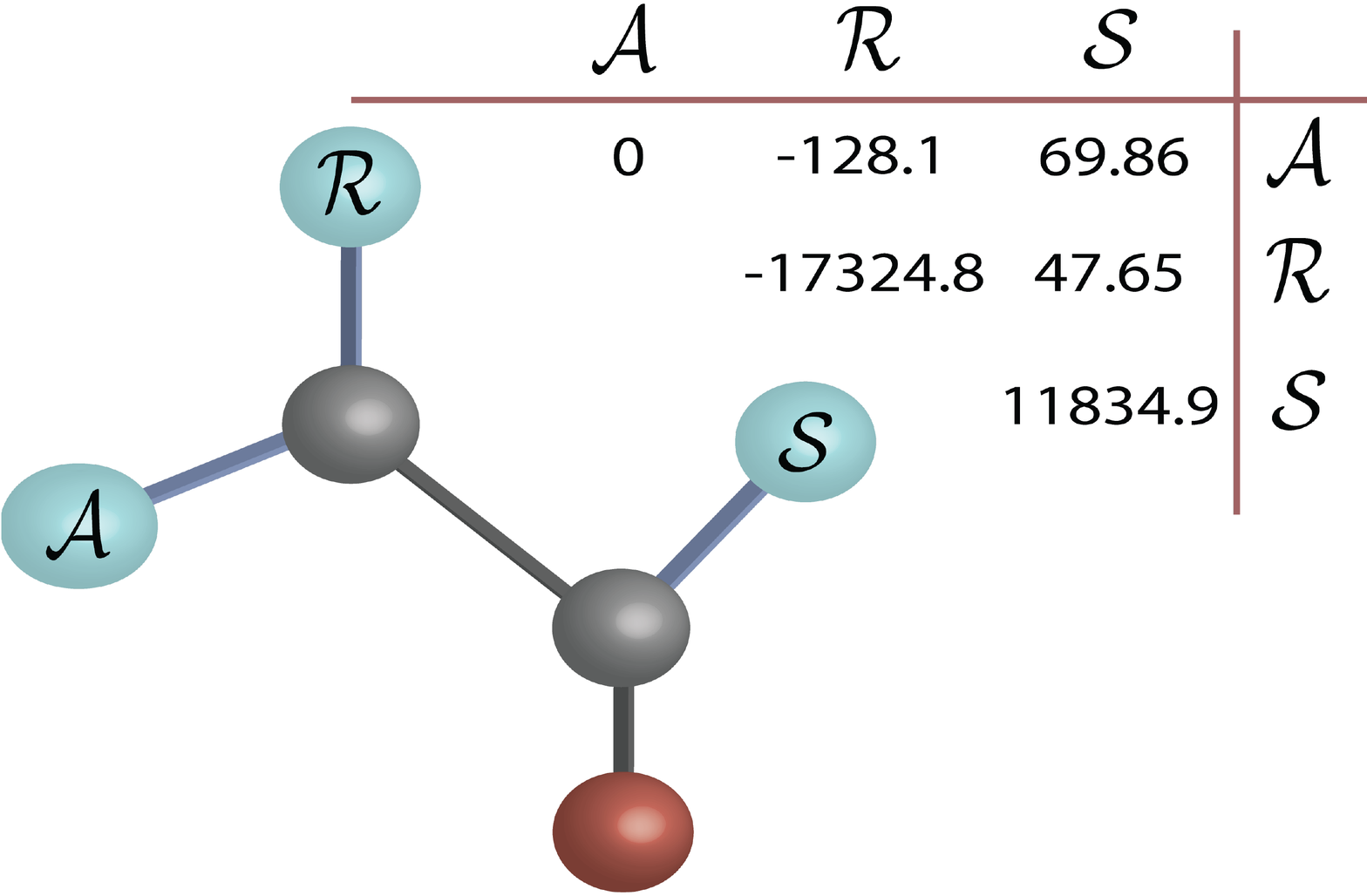}
\end{center}
\caption{{\bf The system.} The molecular structure of the Trifluoroiodoethylene molecule is shown together with its Hamiltonian parameters. The diagonal elements are the relative frequencies, with respect to the ancilla $(\omega_{j} - \omega_{\mathcal{A}})/2\pi$, while the off-diagonal ones are the coupling strengths $J_{i,j}/2\pi$ (see Eq.~\eqref{hamil}). The longitudinal ($T_{1}$) and transversal ($T_{2}^{*}$) relaxation times are also shown. All the frequencies are measured in Hz. Our three qubits are the fluorine nuclei labelled as $\mathcal{A}$, $\mathcal{R}$ and $\mathcal{S}$. The grey spheres represents carbon nuclei while the red on is the iodine. As outlined in \cite{John} by employing the ancilla system we perform the appropriate phase estimation as illustrated in Fig.~\ref{fig:circ}.}
\label{fig:mol}
\end{figure}

\section{Experimental Setup}
In a pulsed NMR experiment a transient signal, called Free Induction Decay (FID) is detected in a pickup coil, following the application of a sequence of radio-frequency pulses. After amplification, this signal is digitised and filtered, before exhibition in a control monitor. The Fourier transform of the FID is the NMR spectrum. The total sample magnetization is proportional to the spectral area, being the proportionality factor dependent on only the electronic circuitry details and resonance frequencies \cite{Raitz}. In the great majority of NMR experiments, however, the detected signal amplitude is understood to be  relative to a reference signal, usually the equilibrium state, and the electronic factor can be neglected. This is the case of the present experiment.

\subsection{The system}

Our experiments were performed using a Varian $500$ MHz Spectrometer with a double resonance probe-head equipped with a magnetic field gradient coil, at room temperature. The sample consists of Trifluoroiodoethylene (C$_{2}$F$_{3}$I) molecules dissolved in acetone D$_{6}$ ($97\%$), whose three $^{19}F$ nuclear spins (spin-$1/2$) represent our qubits (the system, the reservoir and the ancilla). The Hamiltonian of our system is given by the Ising model
\begin{gather}
\hat H = \sum_{j}\hbar\omega_{j}\hat{I}_{z}^{j} + \sum_{j\neq k}\hbar J_{j,k} \hat{I}_{z}^{j}\otimes \hat{I}_{z}^{k} + \hat H_{rf}(t),
\label{hamil}
\end{gather}
with $\hat{I}_{z}^{j} = \hat{\sigma}_{z}^{j}/2$ being the nuclear spin operator in $z-$direction for $j$-th spin ($\sigma_{z}^{j}$ is the Pauli matrix) whose Larmor frequency is $\omega_{j}$. The summations run over the three qubits named: the ancilla $\anc$, the system $\s$ and the reservoir $\e$. The last term of the Hamiltonian,
\begin{gather}
\hat{H}_{rf}(t) = \hbar\omega(t)\left[\hat{\sigma}_{x}^{j}\cos\left(\omega_{rf}t\right) + \hat{\sigma}_{y}^{j} \sin\left(\omega_{rf}t\right)\right],
\end{gather} 
is the radio-frequency Hamiltonian employed to perform any desired unitary operation on the three qubits, by suitably choosing the parameters $\omega(t)$ and $\omega_{rf}$. The physical parameters of our molecule (relaxation times, natural and interaction frequencies)  are shown in Fig.~\ref{fig:mol}. 

\begin{figure}[t]
\begin{center}
\includegraphics[scale=0.15]{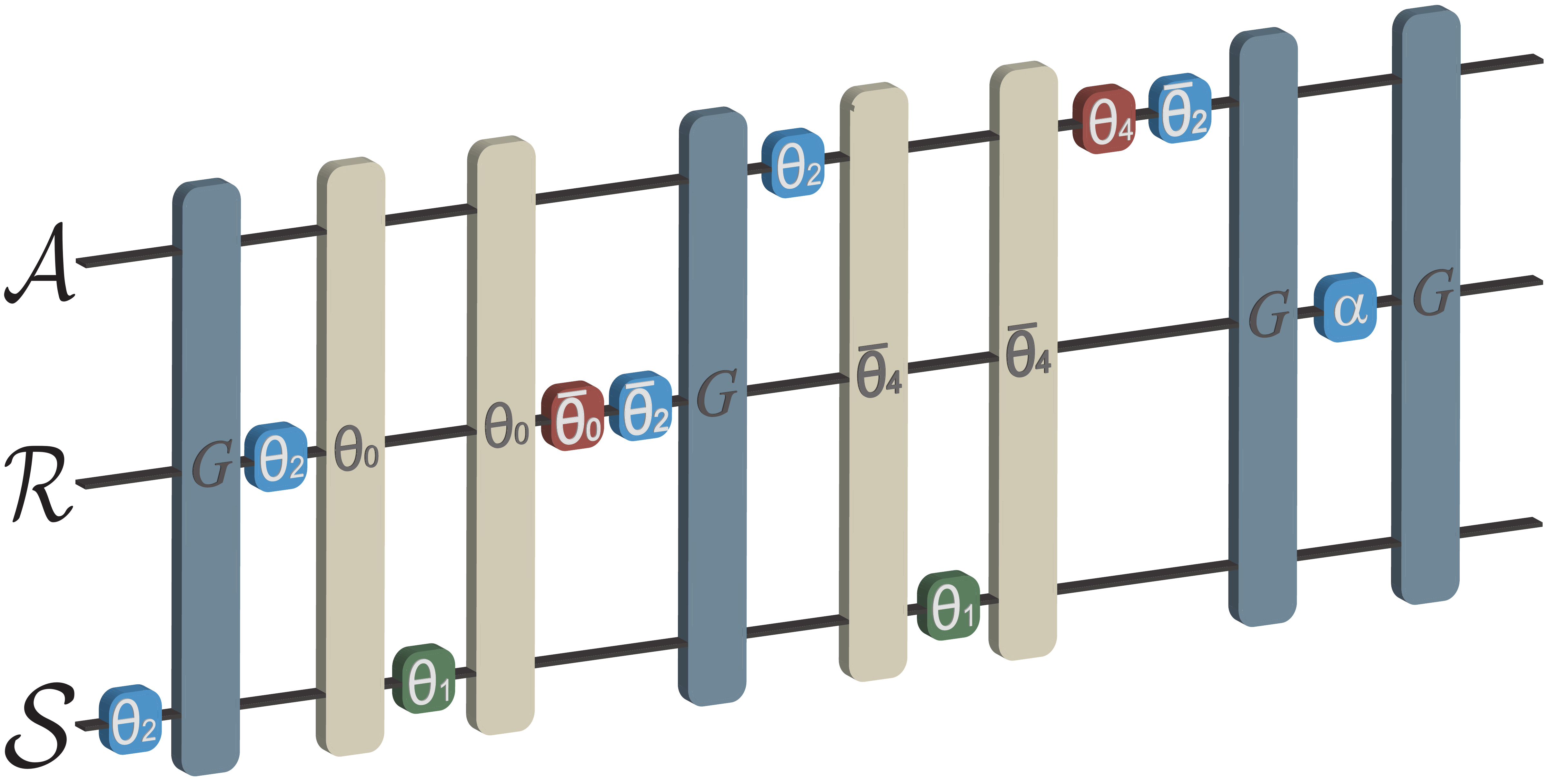}
\end{center}
\caption{{\bf Initial state preparation. }Pulse sequence for the initial state preparation. The green, blue and red squared symbols in each line represents local rotations on the spins over the $x$, $y$ and $z-$directions, respectively. There are two kinds of global operations, the darker ones represent field gradients while the lighter are free evolutions, i.e. evolutions generated by Hamiltonian in Eq.~\eqref{hamil} without the radio-frequency part. The values of the angles inside each symbol, which characterize each operation, are given by $\theta_{0} = -70.5/2^{\circ}$, $\theta_{n} = \pi/n$ for ($n>0$) and $\bar{\theta}_{n} = -\theta_{n}$. The last rotation on the reservoir qubit, denoted by $\alpha$, is used to define the temperature, as explained in the text.}
\label{initial}
\end{figure}

\subsection{Initial state preparation}

In liquid state NMR setup, the system is initially prepared in the so called pseudopure state $\hat{\rho}_{PPS} = (1-\varepsilon) \hat{\mathds{1}}/8 + \varepsilon\hat{\rho}$ instead of $\hat{\rho}$, where $\mathds{1}$ is the identity operator on $\anc\se$ space and $\varepsilon\sim 10^{-5}$ is the ratio between the magnetic and the thermal energy \cite{Gershenfeld,Cory}. Our first goal is to initialize the experiment by preparing $\hat{\rho}_{PPS}$ into the following state 
\begin{gather}
\hat{\rho} = \hat{\rho}_{\mathcal{A}} \otimes \hat{\rho}_{\mathcal{R}} \otimes \hat{\rho}_{\mathcal{S}} = |+\rangle\langle + | \otimes\hat{\rho}_{\mathcal{R}} \otimes \frac{\hat{\mathds{1}}_{\mathcal{S}}}{2}.
\end{gather}
This can be achieved by employing the pulse sequence shown in Fig.~\ref{initial}. In this equation the ancilla qubit, $\mathcal{A}$, is prepared in state $\ket{+}$ (initialised in $\ket{0}$ followed by a Hadamard operation). The reservoir qubit, $\mathcal{R}$, is prepared as
\begin{gather}
\hat{\rho}_{\mathcal{R}} = \left[ 
\begin{array}{cc}
\cos^{2}\left(\frac{\alpha}{2}\right) & 0 \\
0 & 1-\cos^{2}\left(\frac{\alpha}{2}\right)
\end{array}
\right],
\end{gather}
with $\alpha$ being the rotation angle defined in Fig.~\ref{initial}. Comparing this with the definition of the density matrix of a system in thermal equilibrium at finite inverse temperature $\beta$ one can obtain a relation between the temperature and the rotation angle $\alpha$ with the reservoir temperature
\begin{gather}
\beta^{-1} = \frac{2\pi\hbar J_{\mathcal{RA}}}{\log\left[\tan^{2}\left(\frac{\alpha}{2}\right)\right]}.
\end{gather}
From this equation we see that we can prepare states in the range $\beta^{-1} \in [0,\infty)$ by just varying $\alpha$ from $0$ to $\pi/2$. It is important to observe here that this does not correspond to the room temperature, being instead a controlled property of our system that we identify with the temperature of $\e$ by the identification of its state with the Gibbs one, and we can vary this parameter as desired. The system will remain at this temperature for the duration of the experiment if it does not interact with other qubits. Finally, the system, $\mathcal{S}$, is prepared in the maximally mixed state which represents the situation in which it contains one bit of information, thus acting as a memory. Different choices for the system would simply give a different amount of entropy variation and heat dissipated, but the validity of the Landauer's principle is independent of this choice.

\begin{figure}[h]
\begin{center}
\includegraphics[scale=0.15]{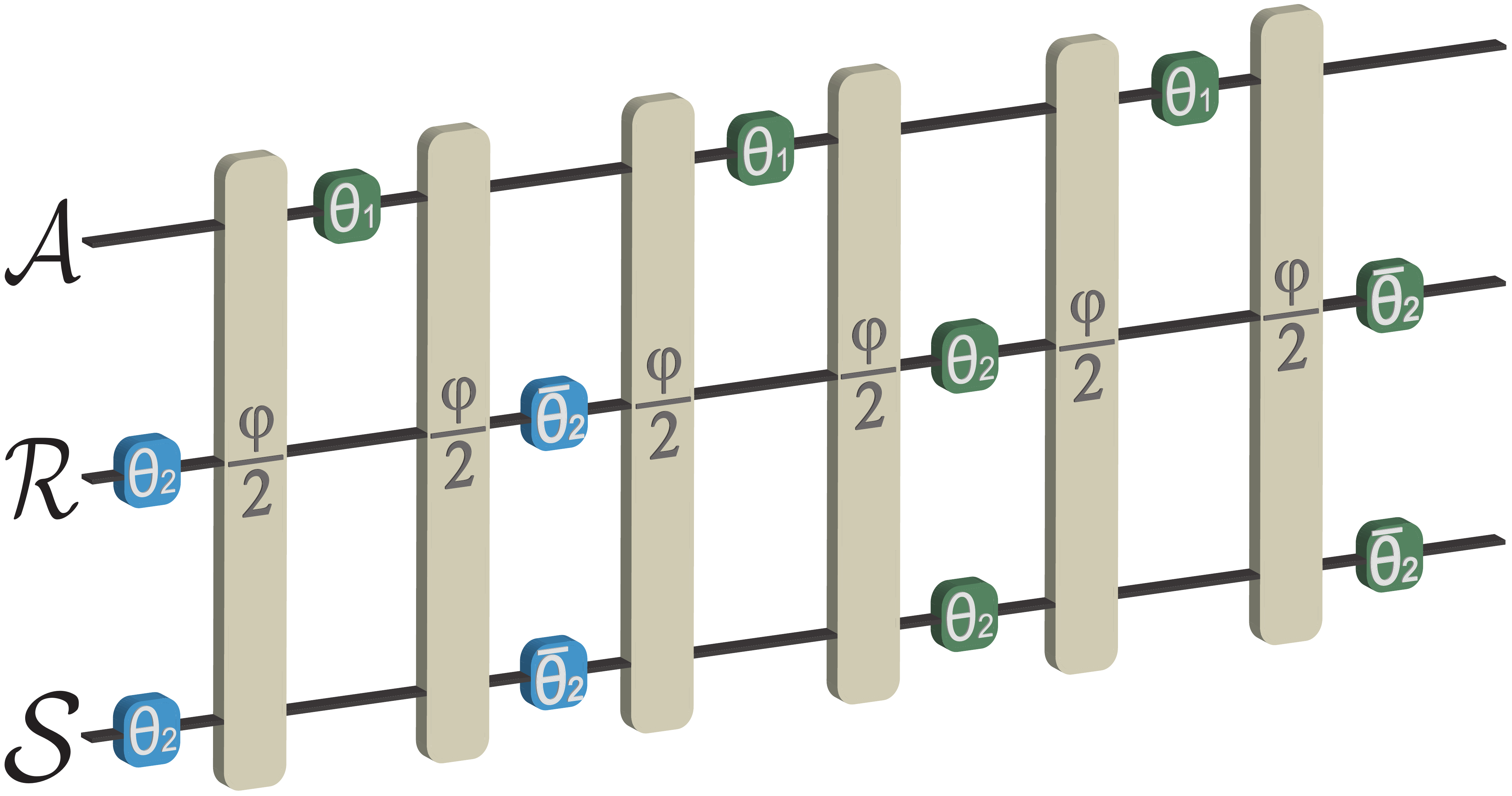}
\end{center}
\caption{{\bf Partial {\sc swap gate}.} Pulse sequence to implement the partial {\sc swap}, $\hat{U}_{PS}(\varphi)$. The labels here follow the same pattern of the ones in Fig.~\ref{initial}. To implement all the required global unitary transformations we need to change the value of $\varphi$, as explained in the text. When $\varphi = 0$ we have the identity operation (on the system and reservoir) while $\varphi = \pi$ and $\varphi = \pi/2$ implements the {\sc swap} and the square root of the {\sc swap} operations (see Eq. \eqref{eqswap}), respectively.}
\label{CircuitU}
\end{figure}

\subsection{The unitary operations}

Unitary operations are implemented by the application of controlled radio frequencies pulses, as shown in Eq.~\eqref{hamil}. Here we investigate Landauer principle considering two distinct processes: $i)$ The partial {\sc swap}, denoted by $\hat{U}_{PS}(\varphi)$ (see Fig.~\ref{CircuitU}) and $ii)$ the controlled-{\sc not}, denoted by $\hat{U}_{CN}$ (see Eq.~\eqref{cnot_s}). The valid of Landauer's principle is completely independent of the choice of these particular operations. Our choice is motivated by the fact that the {\sc Cnot} gate is one of the most employed gates in quantum computation while the {\sc swap} operation mimics the paradigmatic erasure process, usually considered when discussing Landauer's principle and Maxwell's demons. The goal of the next two subsections is to describe the experimental implementations of these operations..

\subsubsection{Partial {\sc swap} operations}

We call partial {\sc swap} the process that continuously interpolates (controlled by the parameter $\varphi$, see Fig.~\ref{CircuitU}) between the identity operation and the total {\sc swap}, passing by the square root of {\sc swap}, which can be represented, in the computational basis, by the matrices
\begin{gather}\label{eqswap}
\hat{U}_{PS}(\pi) = \left[ 
\begin{array}{cccc}
1 & 0 & 0 & 0\\
0 & 0 & 1 & 0\\
0 & 1 & 0 & 0\\
0 & 0 & 0 & 1\\
\end{array}
\right] 
\end{gather}
and
\begin{gather}
\hat{U}_{PS}\left(\frac{\pi}{2}\right) = \left[ 
\begin{array}{cccc}
1 & 0 & 0 & 0\\
0 & \frac{1}{2}(1+i) & \frac{1}{2}(1-i) & 0\\
0 & \frac{1}{2}(1-i) & \frac{1}{2}(1+i) & 0\\
0 & 0 & 0 & 1\\
\end{array}
\right]
\end{gather} 
The square root of {\sc swap} implements half-way swap between the two considered qubits. In our experiment we implemented the operations defined by $\varphi = \lbrace \pi/6, \pi/3, \pi/2, 3\pi/2,5\pi/6, \pi\rbrace$, whose pulse sequences are shown in Fig.~\ref{CircuitU}. The value $\varphi=0$ is the identity operation, while the complete {\sc swap} is implemented by $\varphi = \pi$. This last case represents the complete erasure of the information initially contained in the system, since, after the process, the system ends in the thermal equilibrium state. All other cases can be interpreted as a partial erasure of the system information content. 
 
The main goal of the pulse sequence showed in Fig.~\ref{CircuitU} is to implement a Heinsenberg Hamiltonian between the reservoir and the system, given by   
\begin{eqnarray}
\hat{H}_{h} &=& \hbar(\omega_{\mathcal{R}} - \omega_{\mathcal{A}})\hat{I}_{z}^{\mathcal{R}} + \hbar(\omega_{\mathcal{S}} - \omega_{\mathcal{A}})\hat{I}^{\mathcal{S}}_{z} \nonumber \\
&+& 2\pi\hbar J_{\mathcal{RS}}\left(\hat{I}_{x}^{\mathcal{R}}\otimes \hat{I}_{x}^{\mathcal{S}} + \hat{I}_{y}^{\mathcal{R}}\otimes \hat{I}_{y}^{\mathcal{S}} + \hat{I}_{z}^{\mathcal{R}}\otimes \hat{I}_{z}^{\mathcal{S}}\right).
\label{heisenberg}
\end{eqnarray}
Note that this equation is written in the ancilla rotating frame. This is the chosen reference frame for the experiments performed to determine the characteristic function. Therefore, the effective evolution operator implemented by the pulse sequence in Fig.~\ref{CircuitU} can be written as $\hat{U}_{PS}(\tau) = \exp\{-i \hat{H}_{h} \tau/\hbar\}$. The relation between $\tau$ and the rotation angle $\varphi$ appearing in Fig.~\ref{CircuitU} is given by $\varphi = 2\pi J_{\mathcal{RS}}\tau$, with $J_{\mathcal{RS}} = 47.65 Hz$ in our experiment (see Fig.~\ref{fig:mol}). 

We then vary $\tau$ from $\tau = 0$ (the identity operation) to $\tau = 1/2J_{\mathcal{RS}}$, which implements the complete {\sc swap}. It is important to note here that undesired rotations ---the two first terms in Eq.~\eqref{heisenberg}--- around the $z$ axis are present in our system, but fortunately they can be compensated using the techniques described in Ref.~\cite{Ryan} (see the Appendix).

\subsubsection{Controlled-not operation}

The controlled-{\sc not} gate, which can be represented by the matrix
\begin{gather}
\hat{U}_{CN} = \left[ 
\begin{array}{cccc}
1 & 0 & 0 & 0\\
0 & 1 & 0 & 0\\
0 & 0 & 0 & 1\\
0 & 0 & 1 & 0\\
\end{array}
\right],
\end{gather} 
is a two qubit operation that flips or not the value of the target qubit (in this case the second one) depending on the value of the control qubit (in this case the first one). The pulse sequence for the implementation of this operation is
\begin{gather}
\hat{U}_{CN} = \hat{R}_{x}^{\mathcal{R}}\left(\frac{\pi}{2}\right)\hat{U}_{0}\left(\frac{\pi}{2}\right)\hat{R}_{x}^{\mathcal{A}}\left(\pi\right)\hat{U}_{0}\left(\frac{\pi}{2}\right)\hat{R}_{y}^{\mathcal{R}}\left(\frac{\pi}{2}\right),
\label{cnot_s}
\end{gather}
where $\hat{R}_{k}^{i}\left(\alpha\right)$ is a rotation on the $i-$th qubit about direction $k$ by an angle $\alpha$ while $\hat{U}_{0}$ represents a free evolution, i.e., an evolution generated by Hamiltonian Eq.~\eqref{hamil} without the radio-frequency part. 

\subsubsection{Controlled-$v_t$ operation}

The controlled-$\hat{v}_{t}$ operation, see Fig.~\ref{fig:circ} and Eq.~\eqref{charfunction}, is implemented by letting all of the qubits freely evolve during a time $t/2$ followed by a $\pi$ pulse in the $x-$direction on the system qubit and by another free evolution for $t/2$. The $\pi$ pulse is necessary to effectively protect the system qubit while the desired controlled operation is applied in the other two qubits. For the $\hat{v}_{t}^{\dagger}(t)$ we apply a $\pi/2$ pulse in the $y-$direction on the ancilla qubit, a $t/2$ free evolution, a $\pi$ pulse in the $x-$direction on the system qubit, followed by another $t/2$ free evolution.

\subsection{Measuring the heat distribution and the entropy change}

Our experiment is divided in two parts. First we measure the characteristic function, defined in Eq.~\eqref{charfunction}, by a direct measurement on $\anc$ while varying the reservoir free evolution time $t$ as described in the previous subsection. These gives us the expectation values of $\hat{\sigma}_{x}$ and $\hat{\sigma}_{y}$, which are shown in Fig.~\ref{fig5a}, as a function of time, in one run of the circuit. By computing the discrete inverse Fourier transform of the acquired data for the characteristic function, we attain the corresponding heat distributions, from which we can compute the average dissipated heat, i.e., the left-hand side of Eq.~\eqref{landauer}.

In the second part of the experiment we perform state tomography on $\s$ in order to determine the (average) change in the entropy of the system. See \cite{Ivan} for details on performing quantum state tomography in NMR. From the quantum state tomography data we obtain the quantities appearing in the right-hand side of Eq.~\eqref{landauer}, which characterises the change in information content of the system. In this way we can independently measure both sides of Eq.~\eqref{landauer}, thus verifying Landauer's principle. The results of the experiments are described in the following.

\section{Results}

We have performed two sets of experiments varying the interaction between the system and the environment (the process), one using a {\sc cnot} gate and another one employing the {\sc swap}. In the next two subsections we present the data and corresponding results.

\subsection{{\sc Cnot} gate}

We performed several experiments where the controlled-{\sc not} gate is taken to be the interaction between $\s$ and $\e$. We take $\s$ to be the control qubit and $\e$ to be the target qubit. In these experiments we vary the temperature of the reservoir and the results are shown in Table~\ref{table_data}. As we can see, the measured irreversible entropy production and heat dissipated are in agreement within the errors, confirming Landauer's principle as stated in Eq.~\eqref{landauer}.

\begin{figure}[t]
\begin{center}
\subfigure[] {\includegraphics[width = 250pt]{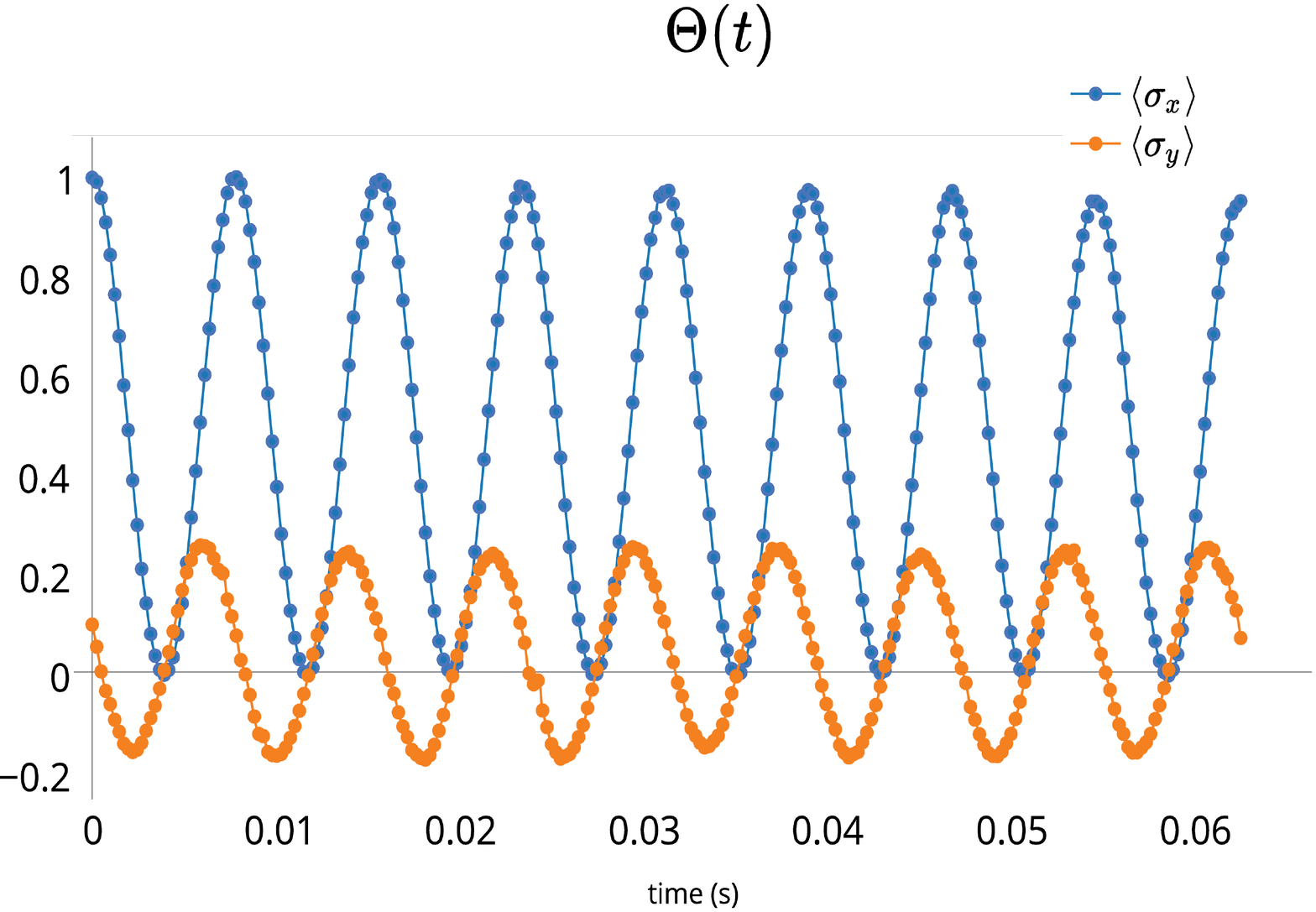}\label{fig5a}}
\subfigure[] {\includegraphics[width = 240pt]{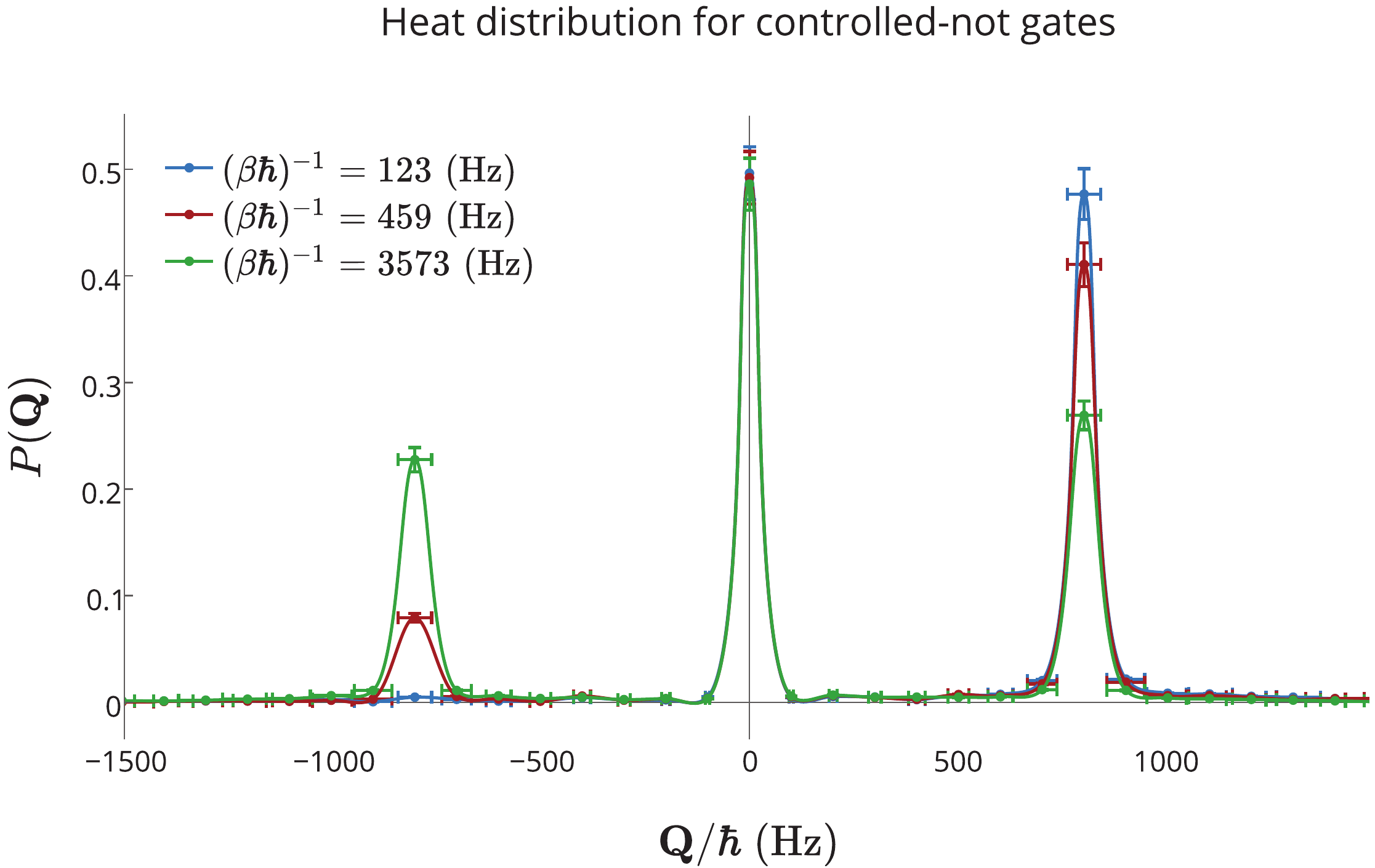}\label{fig5b}}
\end{center}
\caption{{\bf {\sc Cnot} gate experiments.} The first panel shows an example of the measured characteristic function for the {\sc Cnot} gate. The second panel shows the heat distribution, which comes from the discrete Fourier transforming the characteristic function.}
\end{figure}

Figure~\ref{fig5a} shows an example of the experimental characteristic function while in Fig.~\ref{fig5b} we can see examples of the heat distribution at different values of $\beta$. The central peak at $\q=0$, in the heat distribution, corresponds to the cases where the energy eigenstate does not change, while $\q>0$ means a transition from low energy state to high energy state has occurred, and $\q<0$ represents the reverse situation. For this particular gate it is straightforward to see that the theoretical entropy change is $\Delta \ent=0$. However as it is clearly shown, there are instances where $\q < 0$,  seemingly in violation with the Landauer principle. Reinforcing the statistical concept of the second law, these events are fluctuations and the stochastic nature of the thermodynamic variables in this domain is emphasised.  As we can see, although there is a probability to observe a transient \emph{violation} of Landauer's bound in the quantum domain, the \emph{average} heat is greater than the entropy variation, reinforcing the view that Landauer's principle (as well as the second law) are valid on average, but not necessarily for a single realization of an specific experiment. In what follows we now use the average value to explore the heat dissipated for information processing at the ultimate limit.  

\begin{table}
\begin{center}
\begin{tabular}{| c | c | c | c |}
\hline
$ (\beta\hbar)^{-1}$ (Hz) & $\langle\Sigma\rangle\hspace{0.2cm}\mbox{(exp.)}$ & $\beta\langle\mathbf{Q}\rangle\hspace{0.2cm}\mbox{(exp.)}$ & $\Gamma\hspace{0.2cm}\mbox{(theo.)}$ \\ \hline 
123	 & 3.2(2)       & 3.3(2)		& 3.3 \\ \hline 	
185	 & 2.1(1)	     & 2.1(1)       & 2.1 \\ \hline 	
227	 & 1.64(8)	     & 1.66(8)     & 1.67	\\ \hline 
274	 & 1.30(6)	     & 1.32(7)	     & 1.32	\\ \hline 
324	 & 1.03(5)	     & 1.04(5)	     & 1.05\\ \hline 
383	 & 0.80(4)	     & 0.82(4)	     & 0.82\\ \hline 
458	 & 0.61(3)	     & 0.62(3)	     & 0.63\\ \hline 
550	 & 0.45(2)	     & 0.45(2)	     & 0.46\\ \hline 
678	 & 0.31(2)	     & 0.31(2)	     & 0.32\\ \hline 
862   & 0.20(1)	     & 0.20(1)	     & 0.20\\ \hline 
1168 & 0.113(6)	 & 0.114(6)	 & 0.114\\ \hline 
1775 & 0.050(2)	 & 0.052(3)	 & 0.051\\ \hline 
3573 & 0.0128(6)	 & 0.0171(9)	 & 0.0126\\ \hline 
\end{tabular}
\end{center}
\caption{{\bf {\sc Cnot} experiments' data.} Verification of Landauer's theorem for the controlled-not gate for several temperatures. For this case, the theory predicts that $\Delta\mathbf{S} = 0$ (value also obtained experimentally) and Landauer's principle becomes $\langle\Sigma\rangle = I(\hat{\rho}'_{\mathcal{R}}:\hat{\rho}'_\mathcal{S}) + D(\hat{\rho}'_{\mathcal{R}}\|\hat{\rho}_\mathcal{R}) \equiv \beta\langle\mathbf{Q}\rangle$ with $\hat{\rho}' = \hat{U}_{CN}\hat{\rho}\hat{U}_{CN}^{\dagger}$ (see below Eq.~\eqref{land_finite}). As we can see from the data, the Irreversible entropy production due to the implementation of the {\sc Cnot} gate perfectly matches the heat dissipated, both agreeing with the theoretical prediction. The number in parenthesis are the experimental errors, i.e. $3.2(2) = 3.2 \pm 0.2$.}
\label{table_data}
\end{table}

\subsection{{\sc swap} case --- Exploring the Landauer limit}

\begin{figure}[t]
\begin{center}
\subfigure[]{\label{fig3a} \includegraphics[scale=.35]{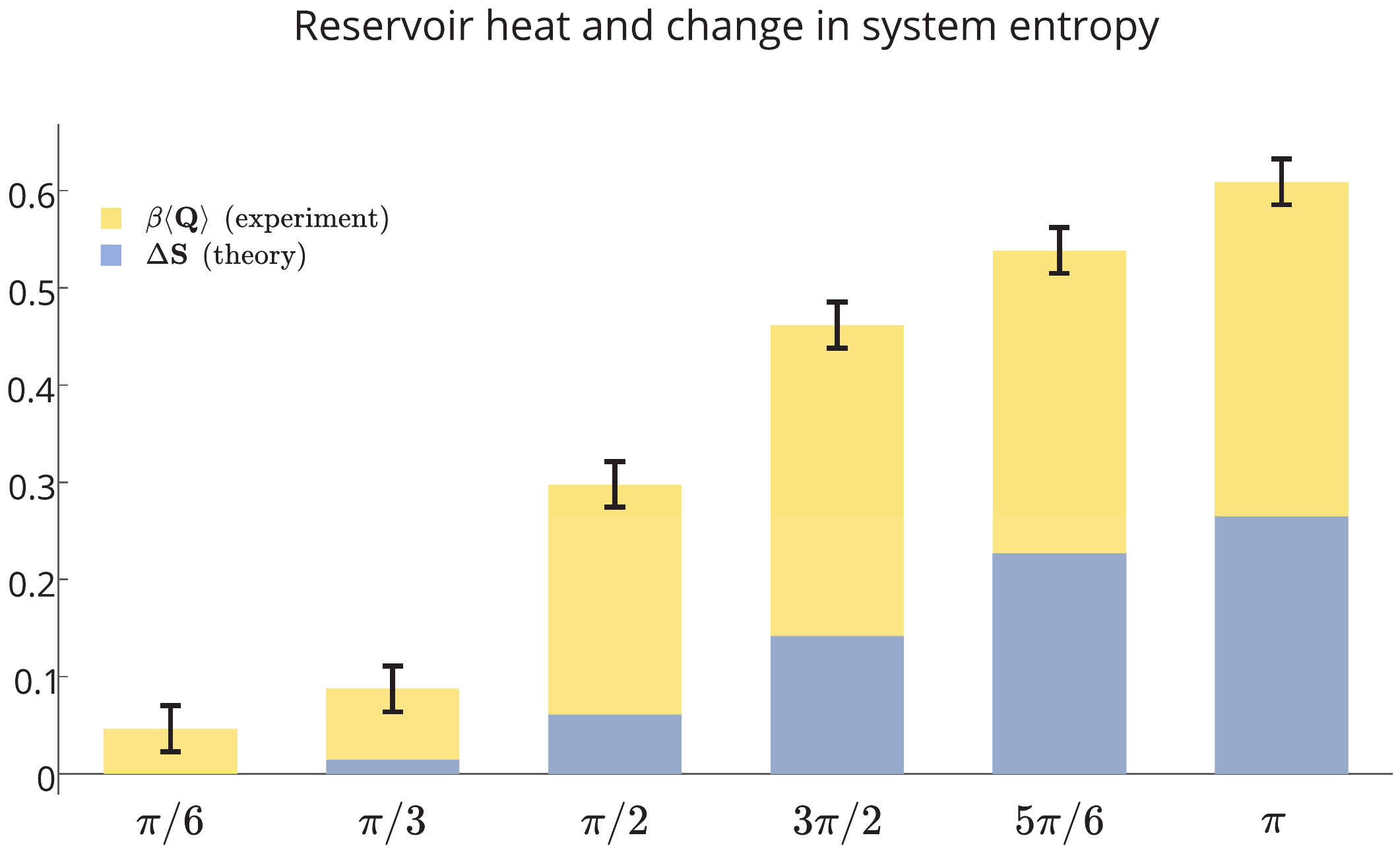}}
\subfigure[]{\label{fig3b} \includegraphics[scale=.35]{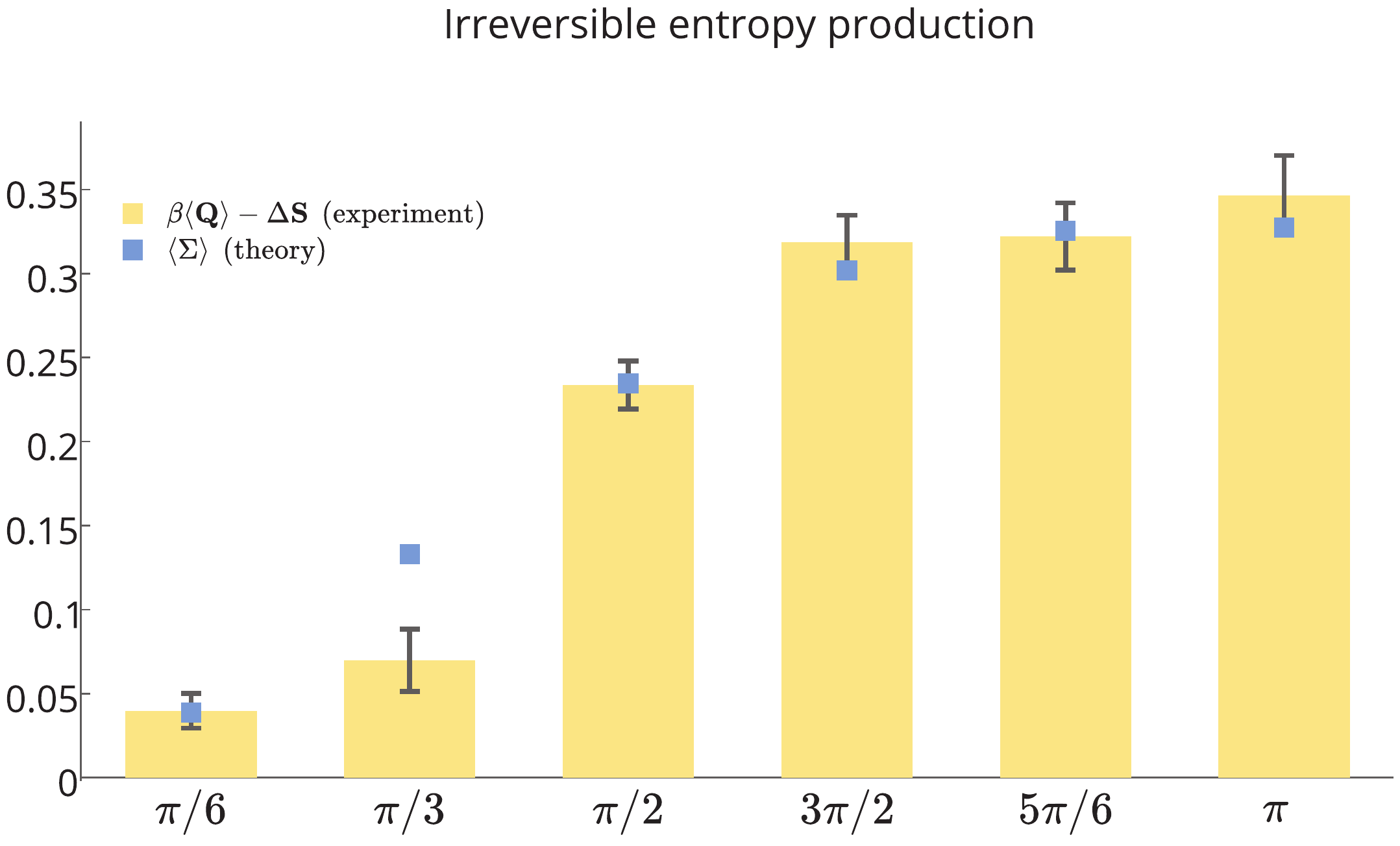}}
\end{center}
\caption{{\bf Landauer limit for partial {\sc swap} gate.} In the first panel we compare the measured average heat generated in $\e$ to theoretical change in entropy of $\s$. The second panel presents the experimentally measured gap between heat and entropy, comparing it to the theoretically computed irreversible entropy production \cite{reeb,esposito1}. Note that both $\Delta \ent$ and $\beta\braket{\q}$ are dimensionless.} 
\end{figure}

In order to reach the Landauer limit we have used the partial {\sc swap} operation, as described earlier. Fig.~\ref{fig3a} shows the average dissipated heat dissipated versus the theoretically computed entropy variation for increasing strength of the process. The case of $\varphi=\pi$ operation which can be seen as the paradigmatic example of the erasure protocol, since the final state of $\s$ is the initial (thermal) state of $\e$, irrespective of the initial state of $\s$. In all cases we confirm that the Landauer principle holds. The feature of Fig.~\ref{fig3a} which initially strikes us is the discrepancy between experiment and theory. This difference is understood as being due to the fundamental irreversible entropy production due to the finite size reservoir. 

It has been  shown by Esposito {\it et al} \cite{esposito1, esposito2} that the average irreversible entropy production $\braket{\Sigma}$ can be computed as 
\begin{gather}
\braket{\Sigma} = \beta \braket{\q} - \Delta \ent,
\label{land_finite}
\end{gather}
The irreversible entropy production has deep meaning in terms of information theory:
$\braket{\Sigma} =\mathcal{I} (\hat\rho'_{\s}: \hat\rho'_{\e}) +\mathcal{D}(\hat\rho'_{\e} \| \hat\rho_{\e})$, where $\mathcal{I}(x:y) := \ent(x) + \ent(y) - \ent(x:y)$ is the mutual information between the system and reservoir at the end of the process and $\mathcal{D} (x \| y) := -\tr[x\log(y)] - \ent(x)$ is the relative entropy between the states of the reservoir before and after the process. The former quantifies the correlations built and the latter the change in the state of the reservoir (see \cite{Nielsen} for details on both of these quantities). It is straightforward to see that in the limit of weak coupling and large reservoir dimension that these terms will vanish and we recover the expected result: $\beta\braket{\q}=\Delta \ent$. It is important to point out that the positivity of the average entropic contribution $\braket{\Sigma}$ was used recently by Reeb and Wolf in order to provide finite size corrections to the Landauer bound \cite{reeb}. In Fig.~\ref{fig3b} we plot the experimentally measured $\braket{\Sigma}$ appearing in the right hand side of the last equation along with the theoretically computed quantity. The agreement between experiment and theory here confirms that we have measured the heat dissipated by an elementary quantum logic gate at the ultimate limit. 

\section{Conclusions}

In this work we have used a modified phase estimation scheme for the extraction of heat statistics from elementary quantum logic gates which were implemented in an NMR experimental setup. The experimental acquisition of the heat statistics allowed us to extract the average heat dissipated during a process at the ultimate limit, set by the Landauer's principle.  Although for the purpose of demonstration we have focused on specific gate implementations, the scheme is sufficiently general to explore entropy production in a range of gate operations and elementary circuits which are central to the theory of quantum information. We believe that the experiments reported in this work will open an avenue for further pioneering experiments on the thermodynamics of systems at the fundamental quantum limit. 

\begin{acknowledgements}
We acknowledge the financial support from the Brazilian funding agencies CNPq (Grants No. 401230/2014-7, 445516/2014-3 and 305086/2013-8), CAPES and the Brazilian National Institute of Science and Technology of Quantum Information (INCT/IQ). This work was partially supported by the COST Action MP1209.
\end{acknowledgements}

\section*{Appendix}

\section*{Error analysis}

To implement single-spin operations, we exploit standard Isech shaped pulses as well as numerically optimized GRAPE pulses \cite{Khaneja}. The GRAPE pulses are optimized to be robust to radio frequency (r.f.) inhomogeneities and chemical shift variations. Two qubit operations were implemented by interleaving free evolutions periods with selective $\pi$ pulses, introduced into the sequences in order to refocalize unwanted evolutions due to the coupling between the spins during the gate implementation.

For combining all operations into a single pulse sequence we have used the techniques described in \cite{Ryan, Bowdrey} for Ising coupled system. A computer program was built, similar to the NMR quantum compiler used in the 7 qubits NMR experiments \cite{Knill,Souza,Zhang}. The imput of this program are the desired unitary transformation, the internal Hamiltonian and predefined shaped pulses. All pulses are then combined  together ensuring that the errors do not propagate as the sequence progresses. The program is capable of minimizing the effects of unwanted coupling evolutions and off-resonance errors as well (see Eq.~\eqref{heisenberg}).

\textbf{Errors in the pulses} --- There are two main error sources here, the signal acquisition (reading) and the duration of each pulse (which is not exactly equal to the planned one). Both these errors were extensively studied in \cite{Raitz}. Assuming that both errors are independent, it was estimated the combined result of $\sim 1\%$ on the measurement of the spins magnetization. However, in order to work with mononuclear systems, shaped pulses are necessary. This improves the precision of the required operations, but increases the duration of the pulses, which contributes to the decoherences processes (see bellow).

\textbf{Errors in the entropies} --- The experimental procedure to determine the entropies requires a smaller amount of pulses (due to the lack of the $\hat{v}_{t}$ operation, which also makes it faster). Therefore, the errors in the entropies are much smaller than the ones in the heat distribution. 

The experimental states were reconstructed by quantum state tomography \cite{Ivan} and the fidelity obtained was, in the worst case, 7\%. The precision of the whole process can be estimated by comparing the fidelities of the measured density matrices and the theoretically calculated ones. For the tested cases we have determined that it was between 2\% (fidelity of $\sim 0.98$) and 7\% (fidelity of $\sim 0.93$), at most. From this, it was possible to estimate, through standard statistical methods, the error bars for the entropies determined on the experiments reported on this paper.

\textbf{Errors in the heat} --- The errors in the heat distribution are caused mainly by two sources. The first one is decoherence, which is discussed bellow. The second one, much more seriously because we cannot correct it, is due to the numerical computation of the inverse Fourier transform of the acquired data. For the determination of the characteristic function only one qubit is measured, which is equivalent to the measurement of the nuclear spin magnetization. This measurement can be achieved with high precision in NMR systems. The error bar for the experimental determination of the average heat was estimated from the standard deviation of the measured points for the characteristic function. Then, we have used standard error propagation for calculating the error bars. In some cases, the oscillations of characteristic function over time are small when compared with the average of $\Theta\left(t\right)$, over time. When this happens we have a larger uncertainty.

\textbf{Decoherence} --- The data acquisition time for the {\sc swap} case varies appreciably, reaching around 150 ms. This is relatively long when compared with the transversal relaxation time for our sample (see Fig.~\ref{fig:mol}). Therefore, the signal lost due to decoherence is considerable and we need to take it into account. To do this we performed numerical simulations of the experiment considering the action of local phase damping channels, which is a very good model for the kind of noise we have \cite{Nielsen}. This noise will cause an exponential decay in the oscillations of the magnetization, whose inverse Fourier transform will give us the heat distribution. The small discrepancies between the simulation and the experiment are mainly due to unwanted couplings not refocused and the inhomogeneity of the radio frequency fields. The net effect was to produce a constant shift in the data both for the heat distribution and for the entropies. We then employed this analysis to correct the final data for signal loss, leading us to the results presented in this work. The controlled-not gates are much faster and the signal loss was not significant. The spin-lattice relaxation, which is characterized by $T_{1}$, causes no appreciable effect during the experiment for both gates.

Therefore, the high level of precision of our setup guarantees that the experimentally implemented operations ({\sc Cnot} and {\sc swap} gates) are very close to the ideal ones, as also confirmed by the excellent agreement between experiment and theory observed here.

\end{document}